\documentclass[authoryear]{elsarticle}

\usepackage{amssymb}

\journal{}

\usepackage{tabularx}
\usepackage{booktabs}
\usepackage{ctable}
\usepackage{hyperref}
\usepackage{url}
\usepackage{listings}
\begin{document}

\begin{frontmatter}
\title{City versus wetland: Predicting urban growth in the Vecht area with a cellular automaton model}


\author[vuearth]{M.~Tend\"{u}r\"{u}s}
\author[cth]{A.G.~Baydin\corref{cor1}}
\ead{atilim@alumni.chalmers.se}
\author[vuenv]{M.A.~Eleveld}
\author[vuenv]{A.J.~Gilbert}
\cortext[cor1]{Corresponding author}
\address[vuearth]{Faculty of Earth and Life Sciences, Vrije Universiteit Amsterdam, De Boelelaan 1085\\1081 HV Amsterdam, The Netherlands}
\address[cth]{Department of Applied Physics, Chalmers University of Technology\\412 96 G\"{o}teborg, Sweden}
\address[vuenv]{Institute of Environmental Studies, Vrije Universiteit Amsterdam, De Boelelaan 1087\\1081 HV Amsterdam, The Netherlands}

\begin{abstract}
There are many studies dealing with the protection or restoration of wetlands and the sustainable economic growth of cities as separate subjects. This study investigates the conflict between the two in an area where city growth is threatening a protected wetland area. We develop a stochastic cellular automaton model for urban growth and apply it to the Vecht area surrounding the city of Hilversum in the Netherlands, using topographic maps covering the past 150 years. We investigate the dependence of the urban growth pattern on the values associated with the protected wetland and other types of landscape surrounding the city. The conflict between city growth and wetland protection is projected to occur before 2035, assuming full protection of the wetland. Our results also show that a milder protection policy, allowing some of the wetland to be sacrificed, could be beneficial for maintaining other valuable landscapes. This insight would be difficult to achieve by other analytical means. We conclude that even slight changes in usage priorities of landscapes can significantly affect the landscape distribution in near future. Our results also point to the importance of a protection policy to take the value of surrounding landscapes and the dynamic nature of urban areas into account.
\end{abstract}

\begin{keyword}
Urban growth \sep stochastic modeling \sep cellular automata \sep wetland \sep landscape protection

\MSC[2010] 91D10 \sep 68U20 \sep 68Q80 \sep 91B72 \sep 91B70
\end{keyword}

\end{frontmatter}

\section{Introduction}
Landscape is defined, in the European Landscape Convention of the \citet{EC2000}, as ``a zone or area as perceived by local people or visitors, whose visual features and character are the result of the action of natural and/or cultural (that is, human) factors''. While human activity is one of the actors in the formation of landscapes, it is also a cause of their destruction. Today, there are unfortunately many landscapes throughout the world that are under serious threat due to factors such as intense use of land, pollution, and insufficient regional planning.

Wetlands, the low-lying marshy swampy lands, are a type of landscape under direct threat by the current pattern of human development. Even though wetlands have been historically considered as a type of wasteland, their importance as integral parts of ecosystems has become increasingly realized during the last few decades \citep{Whigham1992}. Now it is known that they strongly support wildlife habitats, improve water quality, buffer storms, and control erosion and flooding \citep{Kusler1990}. Besides their high natural significance, they also have a high economic value as a natural resource and as areas for recreation. In most developed countries, there are currently many environmental management policies in effect to protect wetlands, while trying to permit their utilization as a natural resource to continue.

\begin{figure}
\centering
\includegraphics[width=0.75\textwidth]{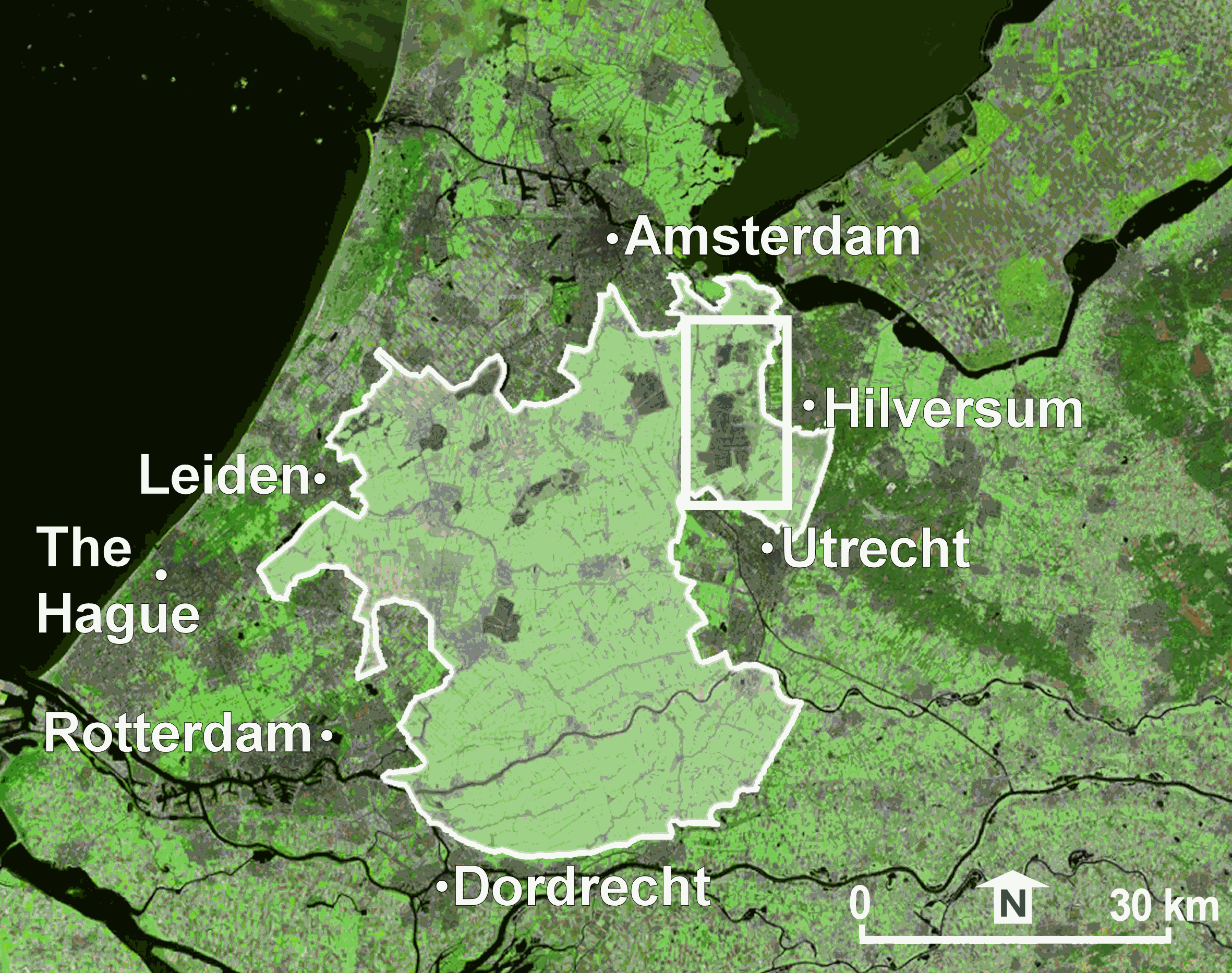}
\caption{Overview of the Vecht area (rectangle), the nearby city of Hilversum, other major cities of the \emph{Randstad conurbation}, and the \emph{Green Heart} of the Netherlands (shaded area in the middle) (Image from Landsat 7. Source: USGS EarthExplorer).}
\label{FigureLocation}
\end{figure}

In this study we intend to analyze a situation where a developing urban area lies very close to a wetland under protection. Urban growth is practically inevitable in a sustainable economy \citep{Henderson2005}. But can a city resist growing to protect a wetland? When will the city threaten the wetland or when will the wetland threaten the city? For how long will it be feasible to protect the wetland? We investigate and discuss these questions here with a case study of the Vecht area in the Netherlands, home to an internationally renowned wetland, and the city of Hilversum lying at the border of the protected area (Fig.~\ref{FigureLocation}).

Our main aim is to investigate the conflicting dynamics of continuing urban expansion and the protection of the local wetland. We present urban growth projections, based on the current distribution of landscapes and their associated importance in terms of natural and economic value, using a stochastic cellular automaton (CA) model. The model produces quantitative and visual information on the changing distribution of landscapes in the Vecht area for the near future (up to year 2055) and is used to predict the anticipated date of clash between the growing city and the wetland. We also study outcomes of several different scenarios with different protection priorities. We believe that insights from this and similar models can be significantly valuable in short and long term regional planning works.

We provide a brief description of the area in our case study in Section~\ref{SectionStudyArea}. This is followed by the details of the method, employed computational model, and data sources in Section~\ref{SectionMethod}, the results of three different runs with the model in Section~\ref{SectionResults}, and the conclusions drawn in Section~\ref{SectionConclusions}.

\section{The study area}
\label{SectionStudyArea}
\subsection{The Vecht area and its urban development policy}

The Vecht area is the floodplain of the river Vecht, forming an important part of the delta river system of the Netherlands. It covers a roughly rectangular area of about 8 km by 20 km and includes wetlands, small lakes, streams, and fen-grassland patches, which create a characteristic flora and fauna with very high biological diversity.

The history of the area through the last centuries has been a unique combination of natural and economic processes: early on dominated by peat extraction forming small ponds, transformation of these into larger lakes via erosion, and the later recovery of the lost land via drainage and regulation of water tables. The area still has a dynamic structure in terms of economy and nature, now being trapped in the middle of the \emph{Randstad conurbation} and considered an important part of the National Ecological Network designated by the Netherlands Environmental Assessment Agency.

The Randstad (also known as Deltametropool) is a conurbation in the western Netherlands where an increased agglomeration of large cities is underway. It is formed mainly by the cities of Amsterdam, Leiden, the Hague, Rotterdam, Utrecht, and Hilversum. These cities surround the Green Heart of the Netherlands, a major open space, of which the Vecht area is a part \citep{VanEck2005,IDG1997}. Fig.~\ref{FigureLocation} gives an overview of the location of the Vecht area relative to the Green Heart and the Randstad. While the Green Heart provides a valuable natural environment and an attractive leisure area for the population, the adjacent cities create new jobs and ask for growing urban development.

The entire Vecht area is also a part of the ``wet axis'' of the National Ecological Network outlined by the Dutch Nature Policy Plan \citep{LNV1990} that aims to connect fragmented natural habitats to improve the region's ecology. This work, in turn, forms a part of a larger plan, Natura 2000, of the European Union for the protection of seriously threatened habitats and the establishment of special protection areas throughout Europe \citep{LNV2005}.

The simultaneous high demand for urban development and nature preservation make the task of spatial planning of urban areas in the region highly challenging. Although there are planning decisions applied strictly to keep the environmental quality high and the development sustainable, progress is yet insufficient as noted in the yearly VINEX reports of the Ministry of Public Housing, Spatial Planning and Environment \citep{VROM2003}. VINEX states that there is a lack of communication and cooperation between municipalities (dealing with construction works) and provinces (dealing with the connection of natural habitats).

\subsection{The city of Hilversum}

\begin{figure}
\centering
\includegraphics[width=0.75\textwidth]{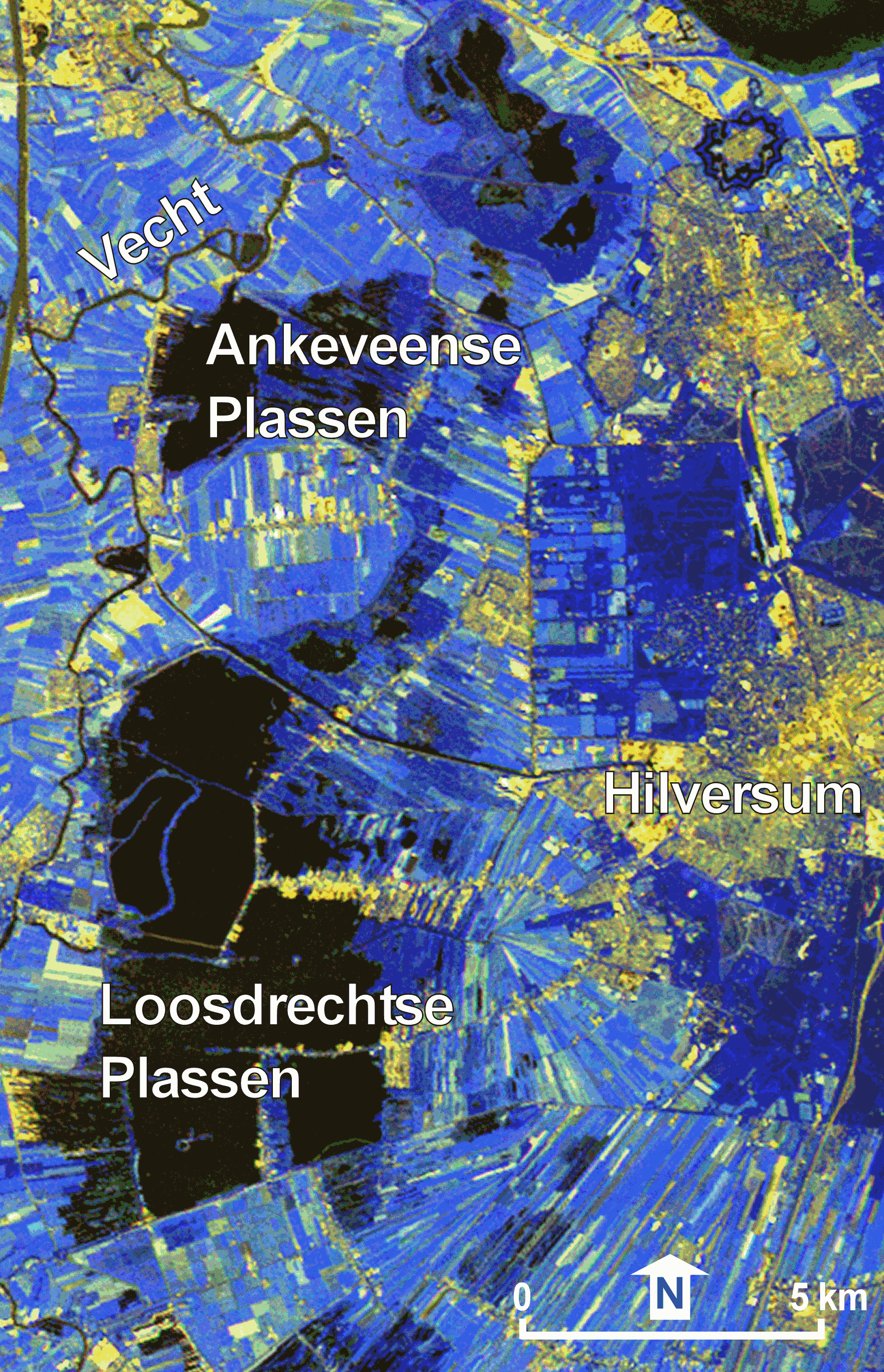}
\caption{The Vecht area and its immediate neighbor, Hilversum (Landsat 7 pseudocolor from bands 1-2-4. Source: USGS EarthExplorer).}
\label{FigureHilversum}
\end{figure}

Since Hilversum is located so close to the Vecht area (Fig.~\ref{FigureHilversum}), it is immediately surrounded by valuable landscapes, comprising not only natural wetlands and wetlands arranged as recreation areas, but also the Cornebos Forest, which is an old forest attracting rare birds, and the heathland of Westerheide \citep{VVV2006}. These areas are highly important in terms of their ecological value. Arable lands and pastures nearby also possess economic and ecological importance. Pasturelands, to some extent, form a transition zone between the wetland and urban areas and provide nesting areas for many bird species.

\begin{figure}[ht]
\centering
\includegraphics[width=70mm]{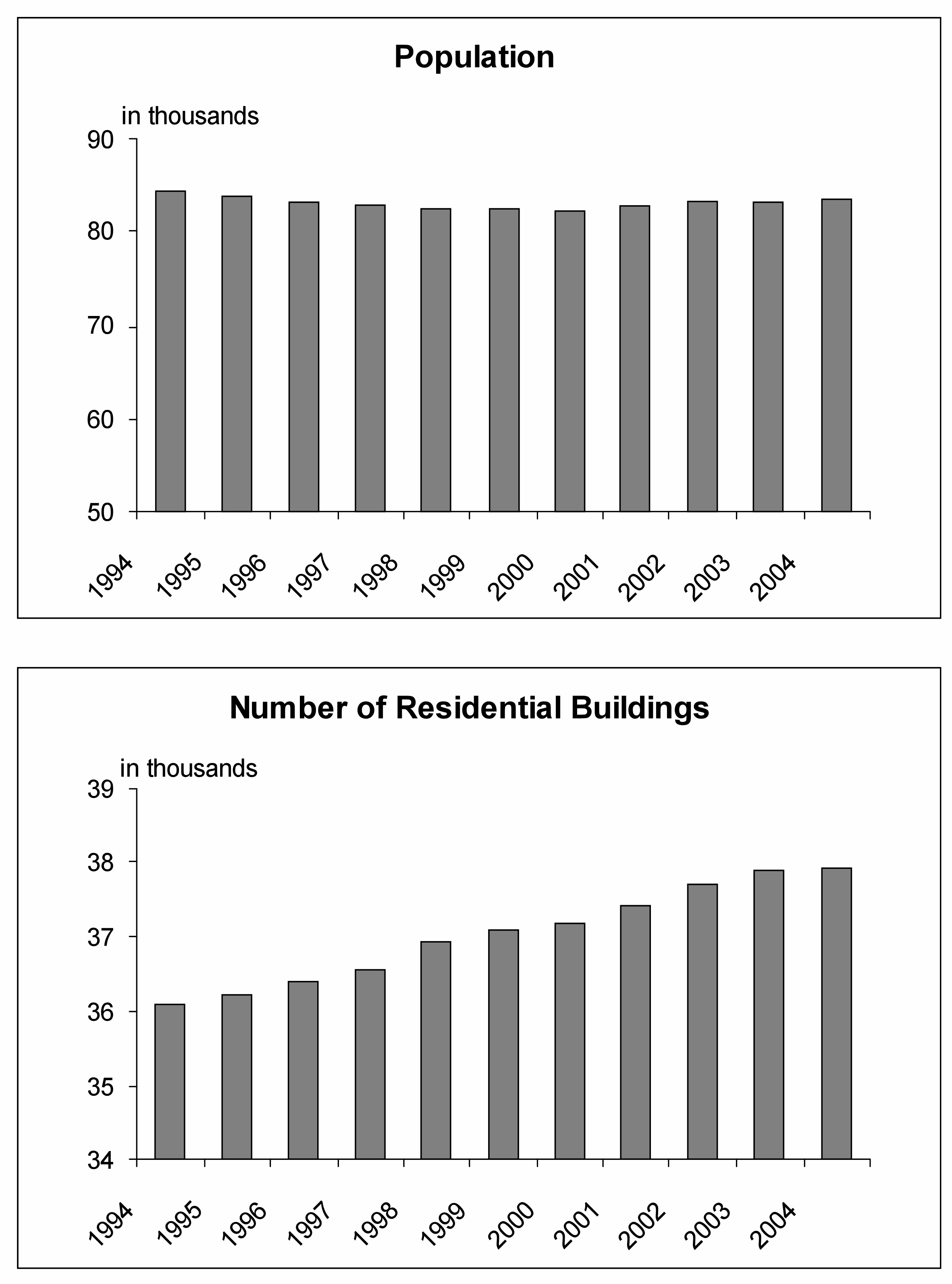}
\caption{Change in the population and the number of residential buildings in the city of Hilversum during the last decade. Source: Statistics Netherlands - CBS StatLine (available online: http://www.cbs.nl/en-GB/menu/cijfers/statline)}
\label{FigureHilversumStatistics}
\end{figure}

Because of the complex structure of the surrounding landscape, city planning in Hilversum is strict. The city has managed to maintain a stable population for the last 10 years to save the protected landscape from harm, but despite low population growth, the extent of the city keeps growing. The number of new housings in a region is suggested as an alternative measure of city growth by \citet{Glaeser1994}. Fig.~\ref{FigureHilversumStatistics} shows the data published by Statistics Netherlands (Het Centraal Bureau voor de Statistiek - CBS), indicating a clear trend of increase in the number of residential buildings, even though the population remains more or less steady during the same period. This suggests that the phenomenon commonly referred to as ``urban sprawl'' is in effect in the area, confirming \citet{Kasanko2006}.

\section{Method}
\label{SectionMethod}

We develop a stochastic cellular automaton (CA) model for urban growth, which is built upon the concept of a \emph{total importance value}\footnote{A simplistic metric summarizing the economic, natural and other values associated with a piece of land.} for each landscape onto which the city may grow. The model is kept as simple as possible at this stage to examine growth dynamics with clarity. We implement the model in C\# programming language and provide the core source code in the appendix. The CA model is further described in Section~\ref{SubSectionModel}.

\subsection{Input and assumptions}

For making projections with the model, we first need to adopt a reasonable average annual rate of urban growth. We assume, by investigating historical maps of the Vecht area, one can get a fairly accurate projection of the middle-term growth rate, provided that the maps are sufficiently detailed. For this purpose, we analyze a collection of topographic maps of the region covering a period between 1850 and 1951, with the majority belonging to the 20th century (Table~\ref{TableMaps}). The reason for not including more recent maps is that during most of this selected period there were no restrictions in place to protect natural habitats, and as such, the measured growth rate would reflect unrestrained factors affecting the growth. This is favorable because it would reproduce the unbounded growth pattern of the city, while the limiting pressure of landscape protection will be dynamically imposed by the CA model in each simulation step on top of this basis growth rate.

For our measurements, we define the area of the city of Hilversum on each historical map by supervised classification using RSI-ENVI 4.1 software. We conclude that an annual urban growth rate of around 2\% is a reasonable value to assume, after observing trends of growth in historical data and giving more weight to the trend during the latter half of the 20th century. This value is also checked against the short-term projected economic growth rate by Statistics Netherlands and is consistent.

For defining the boundaries of the protected wetland area in our model, we employ a suggested restoration plan provided by \citet{VanDenBergh2001} and \citet{Gilbert2004}, which they use to analyze and evaluate land-use scenarios for wetland areas in the Netherlands, but just taking the present situation into account.

\ctable[caption = {The list of employed spatial data.}, label = {TableMaps}, width = {\textwidth}]
{Xp{0.32\textwidth}p{0.15\textwidth}p{0.14\textwidth}}
{
\tnote[a]{Provided by \citet{Geudeke1990}.}
\tnote[b]{Accessible online (\url{https://zulu.ssc.nasa.gov/mrsid}) as compressed MrSID\texttrademark color imagery.}
 }
{\FL
Data & Source & Scale / \linebreak resolution & Date
\ML
Topographic map of Amsterdam and Hilversum & Dienst der Militaire Verkenningen\tmark[a] & 1 : 50 000 & 1850\\
\NN
Topographic map & Topografische Inrichting & 1 : 50 000 & 1909, 1913, 1924, 1931 combined\\
\NN
Topographic map of Utrecht and Amersfoort & Netherlands Ministry of War, Topographic Service & 1 : 50 000 & 1951\\
\NN
Satellite image & GeoCover\texttrademark 2000 -- Landsat 7 ETM+ NASA\tmark[b] & 14.25 m/pixel & c. 2000\\
\NN
Vecht area restoration plan & \citet{VanDenBergh2001} and \citet{Gilbert2004} & N/A & 2004
\LL
}            

\subsection{The model}
\label{SubSectionModel}

While being simple discrete systems of recurrently applied rules, cellular automaton (CA) models have a demonstrated ability to replicate important features of complexity observed in natural processes, as well as having a number of advantages over continuous mathematical models, which are well discussed in CA literature \citep{Schiff2008}. Using CA models for describing urban growth has been receiving increasing attention in the field of urban planning to study the effects of growth and to make predictions in diverse geographic settings \citep{Itami1994,Couclelis1997,Torrens2001}. A well-known CA model of this kind is the SLEUTH (Slope, Land cover, Exclusion, Urbanization, Transportation, and Hillshade) model\footnote{\url{http://www.ncgia.ucsb.edu/projects/gig/}} developed by the United States Geological Survey \citep{USGS2003}, incorporating factors such as the effect of transportation networks and slopes, in addition to the spreading of the existing urban area \citep{Jantz2003}.

Here, we believe that a simpler model will be sufficient for the purposes of this study. The Vecht area is relatively small compared to the scales urban CA models are frequently applied to and this means that the effect of excluding advanced features in the model, such as \emph{breeding},\footnote{A type of growth in urban CA models, in which a new isolated urban cell is occasionally formed without any adjacent urban cells} will be insignificant. Another important factor is the characteristic flatness of the topography of the Netherlands, where most of the country including the Vecht area is virtually void of any hills or slopes \citep{RDG1979}. This renders factors like \emph{slope resistance}, a very important part of general urban CA models, irrelevant in this case.

In our CA model, we consider the area under study as a grid of cells, where each cell is one of 6 types: urban, wetland, heathland, forest, pasture, or water. All cell types except the water type can be urbanized. For initializing the model with the current distribution of landscapes in the region, we employ a recent satellite image (from the GeoCover\texttrademark 2000 set distributed by NASA based on Landsat 7 ETM+, Table~\ref{TableMaps}) and process it with the supervised classification technique. The 6 cell types are assigned real numbers ranging from 0 to 1 representing their importance, where a value of 0 means no protection at all for the cells of that type and a value of 1 means total protection of cells of that type under all circumstances.

\begin{figure}[ht]
\centering
\includegraphics[width=70mm]{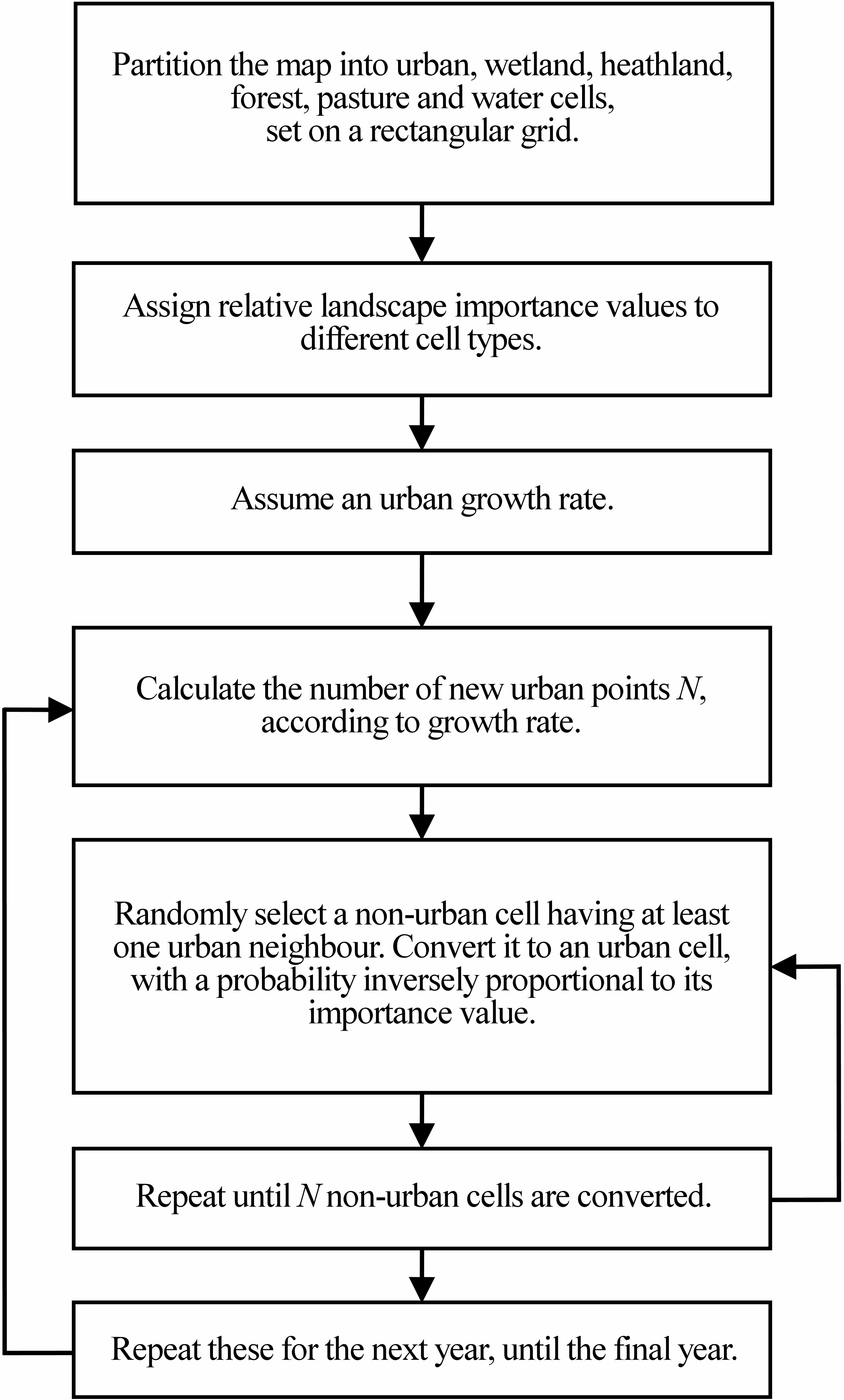}
\caption{Flowchart of the employed urban growth model.}
\label{FigureModel}
\end{figure}

Each year, according to the annual growth rate, a number of new cells are appended to the existing urban area by converting neighboring non-urban cells (i.e. wetland, heathland, forest or pasture) into urban cells. This is done as follows: (1) A random cell is selected on the grid, and then checked for whether it is non-urban or not, and if it has at least one urban neighbor. The random selection continues until such a cell is found. (2) The selected cell is then converted into an urban cell, with a probability inversely proportional to the landscape importance value assigned to its type. (3) For every year during the simulation, this random selection and conversion cycle continues until the required number of new urban cells for that year are created. The model permits the formation of diverse development scenarios by assigning different importance values to different landscape types. The flowchart in Fig.~\ref{FigureModel} gives an outline of the model's working.

The number of new urban cells created each year is simply

\begin{equation}
N' = r \times N
\end{equation}

where $N'$ is the number of new urban cells, $r$ is the annual growth rate, and $N$ is the number of present urban cells. The annual growth rate of urban area is held fixed during a run, based on the assumption that the expected growth rate of a city should be independent of its size \citep{Cordoba2004}.

We take a cautious approach by presuming an ultimate date of reasonable prediction for this model before the year 2055, thinking that calculations covering longer periods would become more diverted by the accumulation of errors introduced by assumptions, selected parameters such as the growth rate, and the intrinsic limitations of the CA modeling approach \citep{Yeh2006}. Also, the trends and priorities in the Vecht area and in the Netherlands in general are likely to be changed beyond terms longer than this.

\section{Results and discussion}
\label{SectionResults}

Going back to the questions at the beginning:

\subsection{Can the city resist growing to protect the wetland?}

This part cannot be answered solely with the aid of the model, but literature and experiences may help. Economic growth is always desired and has a direct influence on the growth of cities. A similar situation described by \citet{Karen2001} in Merced County in the United States, an area that has to deal with urban growth and wetland recreation simultaneously in California's Central Valley, exemplifies how complex the situation could become. It has been noted that \$ 27.7 million per year is spent for the maintenance of the wetland recreation area, and the wetland contributes \$ 41 million per year back to the local economy and provides jobs for more than 800 people. The paradox posed by this situation can immediately be seen: the effort to protect the wetland turns back as a contribution to the local economy, which in turn promotes urban growth and threatens the wetland.

At the same time, wetlands are now recognized globally as a cornerstone and focal point of economic development in both the developed and developing countries and ways are sought after that can balance their protection and use. Wetland restoration in the Vecht area presents many valuable economic benefits to nearby cities, such as income from increased tourism and fishing; and for the city of Hilversum, economic growth is foreseeable. This growth should be planned in a way which ensures the survival of the wetland and the other landscapes in the region.

\subsection{When will the city threaten the wetland or when will the wetland threaten the city?}

To answer this question, we design the first run with the model using parameters thought to represent current policies / priorities in the region. Wetland importance value was set to 1.00 (i.e. it is impossible to destroy wetland) to make sure that it will be protected under all circumstances. Used parameter values are summarized in Table~\ref{TableRun1}.

\begin{figure}[ht]
\centering
\includegraphics[width=\textwidth]{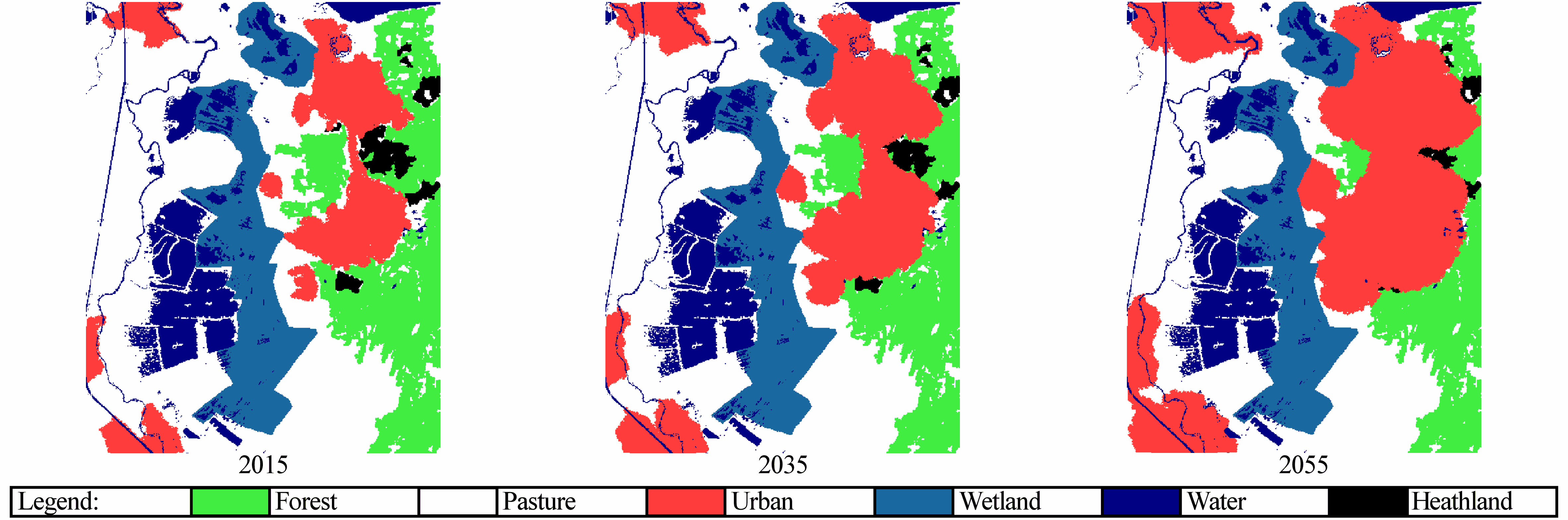}
\caption{The first run of the model with definite protection for the wetland and decreasing importance values for heathland, forest and pasture (details in Table~\ref{TableRun1}). Resulting distribution of landscapes is shown between 2015 and 2055 with 20 year intervals.}
\label{FigureRun1}
\end{figure}

The main result from our simulation, in addition to the projected spatial distribution on the map, is the amount of loss in the area of each landscape type present. Here, the resulting distribution of different landscapes is summarized in Table~\ref{TableRun1} for years 2030 and 2055 (as the amount of area lost and as a ratio of the projected area to the current area) and the projected distribution of landscapes are presented in Fig.~\ref{FigureRun1} from 2015 to 2055 with 20 years interval. Note that the annual urban growth rate during all runs is assumed as 2.0 \%, as explained in Section~\ref{SectionMethod}.

We predict that the clash between the wetland area and the city boundary will become significant by the year 2035, well within a timescale to grant considering in the current recreation and urban plans of the Vecht area.

\ctable[caption = {Parameters and results for the first run. Definite protection for wetland, decreasing importance for heathland, forest and pasture\tmark[a].}, label = {TableRun1}, width = {\textwidth}, pos = {ht}]
{XXXXXX}
{
\tnote[a]{Annual growth rate is 2.0 \%. The simulation starts from year 2005.}
 }
{\FL
Landscape & Importance & \multicolumn{2}{c}{2030} & \multicolumn{2}{c}{2055}
\NN
 & & Amount lost (km$^{2}$) & Remaining (\% of original) & Amount lost (km$^{2}$) & Remaining (\% of original)
\ML
Wetland & 1.00 & 0 & 100.0 & 0 & 100.0
\NN
Heathland & 0.95 & 1.39 & 74.58 & 3.62 & 34.03
\NN
Forest & 0.80 & 5.01 & 88.22 & 14.47 & 66.01
\NN
Pasture & 0.65 & 10.17 & 90.82 & 26.64 & 72.95
\LL
}            

\subsection{How long is this wetland going to be protected?}

The results of the first run in the previous section indicate a deficiency with that scenario. While the wetland is successfully protected, this comes at the cost of losing other valuable landscapes. The most noticeable is the decrease in the total forest area to almost 66\% of its original amount by 2055 (Table~\ref{TableRun1}). One may argue that the proximity of the forest to the urban area makes it more prone to depletion, but it should still be possible to prevent an amount of this loss by trying to adjust the protection priorities.

The other two runs of the model in this section are designed to investigate whether permitting some loss of wetland helps to save a greater amount of the forest. With this purpose, the model was run many times, giving different importance values to each type of landscape and observing the results. Here we present results from two such runs.

\begin{figure}[ht]
\centering
\includegraphics[width=\textwidth]{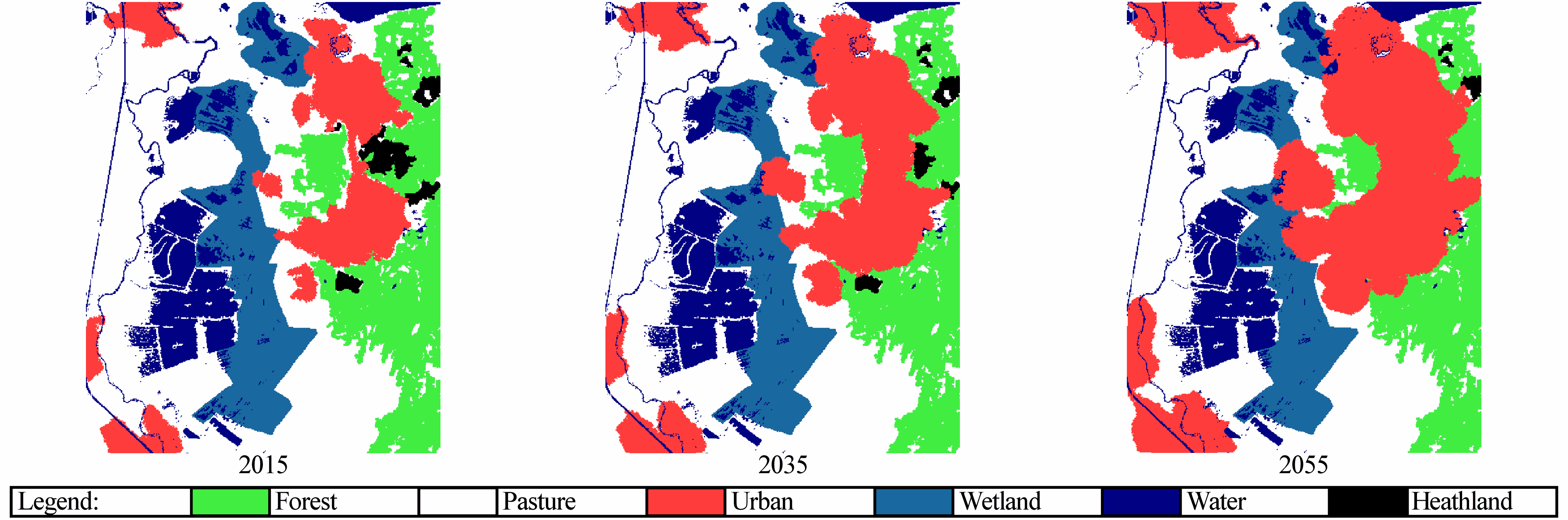}
\caption{The second run of the model, with the wetland importance reduced to 0.90, in an effort to prevent the drastic loss of forest in the first run (details in Table~\ref{TableRun2}). The resulting distribution of landscapes is shown between 2015 and 2055 with 20 year intervals.}
\label{FigureRun2}
\end{figure}

Table~\ref{TableRun2} and Fig.~\ref{FigureRun2} present a situation where the wetland importance parameter is reduced to 0.90, in an effort to prevent the drastic loss of forest in the first run. Surprisingly, it is noticed that with a little sacrifice of the wetland, one has the potential to save a significant amount of forest: around 82\% remaining by 2055, in contrast to the figure of 66\% in the first run (compare Table~\ref{TableRun1} and Table~\ref{TableRun2}).

\ctable[caption = {Parameters and results for the second run. The second run with the model, .}, label = {TableRun2}, width = {\textwidth}, pos = {ht}]
{XXXXXX}
{
\tnote[a]{Annual growth rate is 2.0 \%. The simulation starts from year 2005.}
 }
{\FL
Landscape & Importance & \multicolumn{2}{c}{2030} & \multicolumn{2}{c}{2055}
\NN
 & & Amount lost (km$^{2}$) & Remaining (\% of original) & Amount lost (km$^{2}$) & Remaining (\% of original)
\ML
Wetland & 0.90 & 0.87 & 97.30 & 2.46 & 92.39
\NN
Heathland & 0.90 & 1.69 & 69.27 & 4.35 & 20.68
\NN
Forest & 0.85 & 2.75 & 93.53 & 7.69 & 81.93
\NN
Pasture & 0.65 & 9.13 & 91.76 & 23.55 & 78.75
\LL
}    

\begin{figure}[ht]
\centering
\includegraphics[width=\textwidth]{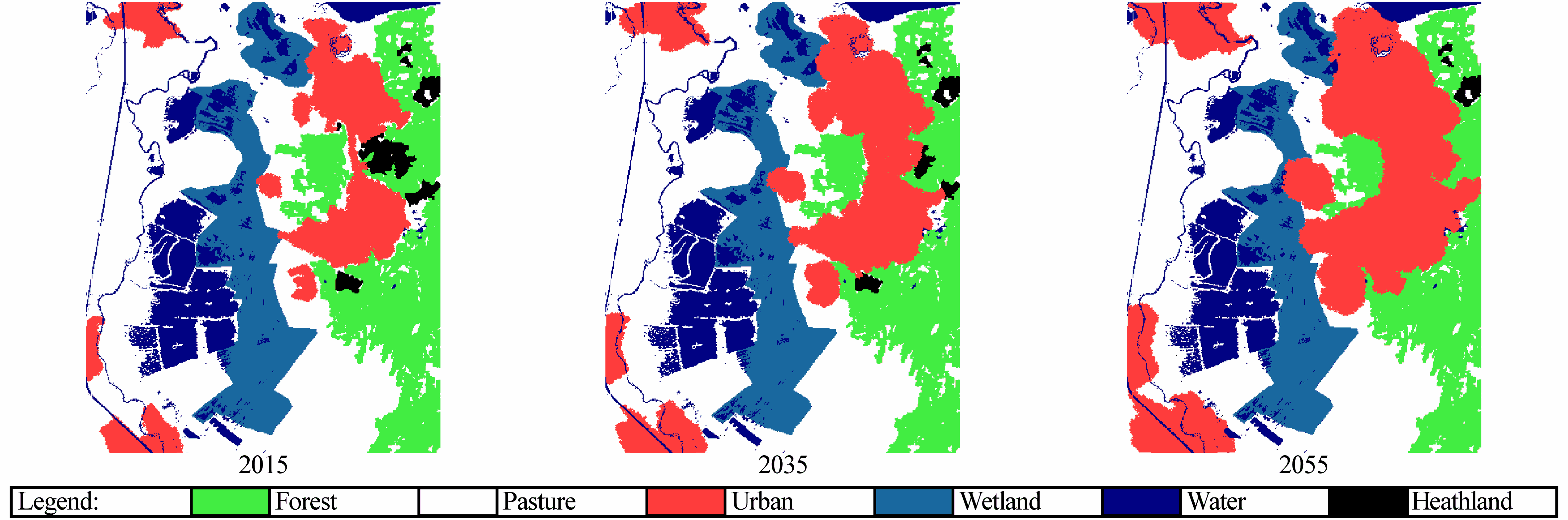}
\caption{The third run of the model with equal importance for the wetland, heathland and forest (details in Table~\ref{TableRun3}). Resulting distribution of landscapes is shown between 2015 and 2055 with 20 year intervals.}
\label{FigureRun3}
\end{figure}

The results of a more radical change in protection policy are presented in Table~\ref{TableRun3} and Fig.~\ref{FigureRun3}, where we assign wetland, heathland and forest equal protection priorities. This results in a situation in which a more even distribution of all resources is achieved while still maintaining the wetland above others. An issue to recognize here is how this is achieved while the wetland is assigned the same importance value as the other landscapes. This is the result of the particular geographical distribution of landscapes in this region: forest and heathland are under immediate threat by the growth of Hilversum, while the wetland lies at some distance, as compared to these.

If wetland protection is considered of highest importance in the Vecht area, then the city of Hilversum will totally grow on other landscapes. In this case, these landscapes will be in danger of irreversible loss to keep the wetland safe. The results presented here indicate that setting a conventional, absolute constraint with the purpose of protecting a type of landscape can lead to other serious problems as a side effect. We conclude that an environmental management project should take the complex interaction and the unique spatial arrangement of different landscapes in a region into account, instead of simply putting particular regions into protected status according to the immediate perceived threat of the day.

\ctable[caption = {Parameters and results for the third run. Equal importance for wetland, heathland and forest\tmark[a].}, label = {TableRun3}, width = {\textwidth}, pos = {ht}]
{XXXXXX}
{
\tnote[a]{Annual growth rate is 2.0 \%. The simulation starts from year 2005.}
 }
{\FL
Landscape & Importance & \multicolumn{2}{c}{2030} & \multicolumn{2}{c}{2055}
\NN
 & & Amount lost (km$^{2}$) & Remaining (\% of original) & Amount lost (km$^{2}$) & Remaining (\% of original)
\ML
Wetland & 0.80 & 1.16 & 96.40 & 3.38 & 89.54
\NN
Heathland & 0.80 & 2.28 & 58.36 & 4.35 & 20.65
\NN
Forest & 0.80 & 2.33 & 94.52 & 6.95 & 83.67
\NN
Pasture & 0.65 & 8.69 & 92.15 & 23.29 & 78.98
\LL
}    

\subsection{Further discussion}

There is another way to protect those landscapes from the threatening urban growth in Hilversum, and it is even more effective: promoting a vertical growth (i.e. higher buildings) instead of the current trend of lateral growth. This solution can be a very efficient in protecting the valuable landscapes in the Vecht area, but there is no indication that this will be happening in the near future.

An important insight gained from this study is on the mode of propagation of a growing city in such a configuration of landscapes and the speed at which this occurs. This type of information can be very useful in environmental management projects by making it easier to pick which urban areas in the proximity of a protected area to get included in the protection plan (by predicting whether an urban area will pose a significant threat during the course of the protection plan).

Another issue of importance is how the suggested protection priorities, as represented in our model by the so-called importance values, could be implemented in reality, while how these priorities should be determined remains a separate subject. The issue often exceeds environmental considerations and involves other factors such as community demand for each landscape type. A recent study by \citet{Kaplan2004} on community preferences about neighboring different landscapes, for instance, has shown that there is a considerably greater desire for forests than other landscapes, regardless any benefits that might be obtained from these.

\section{Conclusions}
\label{SectionConclusions}

Although urban growth is a dynamic and complex process, significant insight can still be gained by experimentation with computational models. An investigation of the Vecht area and the city of Hilversum is particularly useful in demonstrating this modeling and prediction approach, given the diversity of landscapes present in the region and the immediate threat posed on these by urban growth.

We predict that the pressure to protect the designated wetland area will probably keep the city away from the wetland until around 2035, after which the expected clash between the city and the protection area will occur. But up to that time, the city would have to grow into other valuable landscapes, including forests. This leads to our suggestion that, to protect not only the wetland but also the diversity of other landscapes, the protection pressure on the wetland should be reduced to a level bringing the anticipated clash to an earlier date, while still preserving a considerable amount of wetland. This way, we believe, a more even distribution of all resources is achieved while still maintaining the wetland above others. An important aspect to recognize here is how this can be achieved even while the wetland is assigned the same importance value as the other landscapes.

It is certain that landscapes will change continuously. But this change should be monitored and controlled in a way that is most beneficial for nature and human society. By investigating this interplay through experiments with different protection scenarios, this study shows that even a slight adjustment in the utilization priorities of landscapes is capable of achieving important changes in the distribution of landscapes that we will be depending on in the near future.

\section{References}

\bibliographystyle{elsarticle-harv}
\bibliography{TendurusBaydinEleveldGilbert2010}

\newpage
\appendix
\section{Code for the CA model}

Following is the core C\# code of the CA urban growth model used in this study. For brevity, only the portion central to the working of the model is presented, leaving out auxiliary code such as for the interface. The code was compiled and executed on Microsoft .NET Framework version 2.0.50727.

\small
\begin{lstlisting}[breaklines=true]

namespace HilversumVecht
{
    public class CAModel
    {
        private enum CellType {Forest, Heathland, Urban, Wetland, Water, Canal, Void}

        //Predefined RGB values for:
        //1. Reading the cell types from the initial map presented to the class
        //as a bitmap via the constructor (Bitmap initialMap).
        //2. Producing the output bitmap while the model is running.
        private static Color ModelMapForestColor = Color.FromArgb(121, 174, 4);
        private static Color ModelMapHeathlandColor = Color.FromArgb(219, 198, 215);
        private static Color ModelMapUrbanAreaColor = Color.FromArgb(231, 56, 13);
        private static Color ModelMapWetlandColor = Color.FromArgb(13, 111, 55);
        private static Color ModelMapWaterColor = Color.FromArgb(46, 143, 210);
        private static Color ModelMapCanalColor = Color.FromArgb(107, 175, 220);
        private static Color ModelMapVoidColor = Color.FromArgb(195, 212, 144);
        private CellType[,] ModelMap;
        private int Width, Height;
        private float ForestValue, WetlandValue, HeathlandValue, PastureValue;
        private bool Abort;
        private Random rnd;

        // The constructor of the class.
        public CAModel(int width, int height, Bitmap initialMap, float forestValue, float wetlandValue, float heathlandValue, float pastureValue)
        {
            Width = width; Height = height; ForestValue = forestValue; WetlandValue = wetlandValue; HeathlandValue = heathlandValue; PastureValue = pastureValue;
            rnd = new Random();
            ModelMap = ReadModelMap(initialMap);
        }

        // Reads the initial landscape distribution from the bitmap.
        private CellType[,] ReadModelMap(Bitmap map)
        {
            CellType[,] ret = new CellType[Width, Height];
            for (int x = 0 ; x < Width ; x++)
                for (int y = 0 ; y < Height ; y++)
                    ret[x, y] = ColorToCellType(map.GetPixel(x, y));
            return ret;
        }

        // Converts a color in the initial landscape distribution bitmap to a CellType value.
        private static CellType ColorToCellType(Color c)
        {
            if (c == ModelMapForestColor)
                return CellType.Forest;
            else if (c == ModelMapHeathlandColor)
                return CellType.Heathland;
            else if (c == ModelMapWetlandColor)
                return CellType.Wetland;
            else if (c == ModelMapUrbanAreaColor)
                return CellType.Urban;
            else
                return CellType.Void;
        }

        // Converts a CellType value into the corresponding color for producing the bitmap output.
        private static Color CellTypeToColor(CellType c)
        {
            if (c == CellType.Forest)
                return ModelMapForestColor;
            else if (c == CellType.Heathland)
                return ModelMapHeathlandColor;
            else if (c == CellType.Wetland)
                return ModelMapWetlandColor;
            else if (c == CellType.Urban)
                return ModelMapUrbanAreaColor;
            else if (c == CellType.Void)
                return ModelMapVoidColor;
            else
                return Color.Yellow;
        }

        // Main method for running the model.
        public void Run(float growthRate, int steps)
        {
            Abort = false;
            float initialForestArea = CountArea(CellType.Forest);
            float initialWetlandArea = CountArea(CellType.Wetland);
            float initialHeathArea = CountArea(CellType.Heathland);
            float initialUrbanArea = CountArea(CellType.Urban);
            float initialOtherArea = Width * Height - initialUrbanArea - initialWetlandArea - initialForestArea;
            int fa, wa, ha, ua;
            float percentForestArea, percentWetlandArea, percentUrbanArea, percentHeathArea, percentOtherArea;
            Bitmap map = new Bitmap(Width, Height);
            for (int i = 0 ; (i < steps) && !Abort ; i++)
            {
                UpdateModelMap(growthRate);
                fa = CountArea(CellType.Forest); wa = CountArea(CellType.Wetland); ha = CountArea(CellType.Heathland); ua = CountArea(CellType.Urban);
                percentForestArea = (fa / initialForestArea) * 100; percentWetlandArea = (wa / initialWetlandArea) * 100; percentUrbanArea = (ua / initialUrbanArea) * 100; percentHeathArea = (ha / initialHeathArea) * 100; percentOtherArea = ((Width * Height - fa - wa - ua - ha) / initialOtherArea) * 100;
                map = WriteModelMap();
            }
        }

        // The annual update to the grid.
        private void UpdateModelMap(float growthRate)
        {
            int px, py, timeout;
            int steps = (int)((growthRate / 100) * CountArea(CellType.Urban));
            for (int i = 0 ; i < steps ; i++)
            {
                //Timeout introduced for performance concerns
                timeout = 100000;
                while (timeout > 0)
                {
                    px = rnd.Next(1, Width); py = rnd.Next(1, Height);
                    if (ModelMap[px, py] != CellType.Urban)
                    {
                        if (UrbanNeighbors(px, py) > 2)
                        {
                            if (ModelMap[px, py] == CellType.Void)
                            {
                                if (PastureValue <= rnd.NextDouble())
                                {
                                    ModelMap[px, py] = CellType.Urban; break;
                                }
                            }
                            else if (ModelMap[px, py] == CellType.Forest)
                            {
                                if (ForestValue <= rnd.NextDouble())
                                {
                                    ModelMap[px, py] = CellType.Urban; break;
                                }
                            }
                            else if (ModelMap[px, py] == CellType.Wetland)
                            {
                                if (WetlandValue <= rnd.NextDouble())
                                {
                                    ModelMap[px, py] = CellType.Urban; break;
                                }
                            }
                            else
                            {
                                if (HeathlandValue <= rnd.NextDouble())
                                {
                                    ModelMap[px, py] = CellType.Urban; break;
                                }
                            }
                        }
                    }
                    timeout--;
                }
            }
        }

        // Produces the output bitmap from the current grid (variable "ModelMap").
        private Bitmap WriteModelMap()
        {
            Bitmap ret = new Bitmap(Width, Height);
            for (int x = 0 ; x < Width ; x++)
                for (int y = 0 ; y < Height ; y++)
                   ret.SetPixel(x, y, IntegerToColor(ModelMap[x, y]));
            return ret;
        }

        // Counts the number of urban neighbors of a given cell.
        private int UrbanNeighbors(int x, int y)
        {
            //Moore neighborhood
            int ret = 0;
            if (ModelMap[x - 1, y - 1] == CellType.Urban)
                ret++;
            if (ModelMap[x, y - 1] == CellType.Urban)
                ret++;
            if (ModelMap[x + 1, y - 1] == CellType.Urban)
                ret++;
            if (ModelMap[x - 1, y] == CellType.Urban)
                ret++;
            if (ModelMap[x + 1, y] == CellType.Urban)
                ret++;
            if (ModelMap[x - 1, y + 1] == CellType.Urban)
                ret++;
            if (ModelMap[x, y + 1] == CellType.Urban)
                ret++;
            if (ModelMap[x + 1, y + 1] == CellType.Urban)
                ret++;
            return ret;
        }

        // Counts the total number of cells of a given CellType on the grid.
        private int CountArea(CellType c)
        {
            int ret = 0;
            for (int x = 0 ; x < Width ; x++)
                for (int y = 0 ; y < Height ; y++)
                    if (ModelMap[x, y] == c)
                        ret++;
            return ret;
        }
    }
}

\end{lstlisting}

\end{document}